\documentclass[english]{article}
\usepackage[T1]{fontenc}
\usepackage[latin1]{inputenc}
\usepackage{amsmath}
\usepackage{setspace}
\usepackage{amssymb}

\makeatletter
\usepackage{latexsym}

\usepackage{babel}
\makeatother
\begin{document}
\begin{flushright}TUW-05-15\end{flushright}

\bigskip{}
\begin{center}\textbf{\large PROGRESS AND PROBLEMS IN QUANTUM GRAVITY}\end{center}{\large \par}

\bigskip{}
\begin{center}W. Kummer\end{center}

\medskip{}
\begin{center}\textit{Institute f. Theor. Physics, Vienna University
of Technology}\end{center}

\begin{center}\textit{Wiedner Hauptstraße 8-10, A-1040 Vienna, Austria}\end{center}

\begin{abstract}
From the point to view of an uncompromising field theorist quantum
gravity is beset with serious technical and, above all, conceptual
problems with regard especially to the meaning of genuine {}``physical''
observables. This situation is not really improved by the appearance
of recent attempts to reformulate gravity within some novel framework.
However, the original aim, a background independent quantum theory
of gravity can be achieved in a particular area, namely 2d dilaton
quantum gravity without any assumptions beyond standard quantum field
theory. Some important further by-products of the research of the
{}``Vienna School'' include the introduction of the concept of Poisson-Sigma
models, a verification of the {}``virtual Black Hole'' and the extensions
to $N=(1,1)$and $N=(2,2)$ 2d-supergravity, for which complete solutions
of some old problems have been possible which are relevant for superstring
theory.\newpage

\end{abstract}

\section{Introduction}

The conclusion cannot be avoided that a merging of quantum theory
with Einstein's theory of general relativity%
\footnote{Several reviews of quantum gravity have emerged at the turn of the
millennium, cf. e.g. \cite{Hor-1},\cite{Car-2}.%
} (GR) is necessitated by consistency arguments. E.g. the interaction
of a classical gravitational wave with a quantum system inevitably
leads to contradictions \cite{Epp-3}: If that interaction leads to
a collapse of the probability function, momentum conservation breaks
down, if it does not, signals must proceed at velocities larger than
the speed of light.

When a quantum theory (QT) of gravity is developed along usual lines
of quantum field theory (QFT), one is confronted with a fundamental
problem, from which many other (secondary) difficulties can be traced.
The crucial difference to flat space is the fact that the variables
of gravity exhibit a dual role, they are fields living on a manifold
which is determined by themselves, {}``stage'' and {}``actors''
coincide. Any separation between those aspects may even be the origin
of perturbative non-renormalizability \cite{Hoo-4}. There exist further
problems: the time variable, an object with special properties already
in QT, in GR appears on an equal footing with the space coordinates.

In section 2 we shortly recall the definition of physical variables
in QFT and the ensuing ones in quantum gravity (QGR).

Then (section 3) we turn to those quantities which can be interpreted
as {}``physical observables'', both at the classical and at the
quantum level. In the latter case the most serious problems arise
in QGR - and are almost completely ignored in the contemporary literature.

In the last decade several new approaches to GR and QGR have been
introduced which are described in section 4.

Finally (section 5) we mention some highlights of the {}``Vienna
approach'' to 2d dilaton quantum gravity. In that area which contains
also models with physical relevance (e.g. spherically reduced gravity)
the application of just the usual concepts of (even nonperturbative!)
QFT lead to very interesting consequences \cite{Gru-5} which allow
physical interpretations in terms of {}``solid'' traditional QFT
observables.

\section{The Field Variables in GR}

The field variables of a QFT usually are not directly accessible to
experimental measurements. Traditionally the metric $g_{\mu\nu}$
is the one used in GR. However, from a more fundamental geometric
point of view \cite{Nak-6} the metric is a {}``derived'' field
variable

\begin{equation}
g=e^{a}\otimes e^{b}\eta_{ab},\label{1}\end{equation}

\noindent because it is the direct product of the dual basis one forms%
\footnote{For details on gravity in the Cartan formulation we refer to the mathematical
literature, e.g. \cite{Nak-6}.%
} $e^{a}=e_{\mu}^{a}dx^{\mu}$ con tracted with the flat local Lorentz
metric $\eta_{ab}$ which is used to raise and lower {}``flat indices''
denoted by Latin letters $\left(\eta=diag\left(1,\,-1,\,-1,\,-1,....\right),x^{\mu}=\left\{ x^{0},\, x^{i}\right\} \right)$.
Local Lorentz invariance leads to the {}``covariant derivative''
$D_{\, b}^{a}=\delta_{b}^{a}d+\omega_{\,\,\, b}^{a}$ with a spin
connection 1-form $\omega_{\,\,\, b}^{a}$ as a gauge field. Its antisymmetry
$\omega^{ab}=-\omega^{ba}$ implies metricity. Thanks to the Bianchi
identities all covariant tensors relevant for constructing actions
in even dimensions can be expressed in terms of $e^{a},$the curvature
2-form $R^{ab}=\left(D\omega\right)^{ab}$ and the torsion 2-form
$T^{a}=\left(De\right)^{a}.$ For nonvanishing torsion the affine
connection $\Gamma_{\mu\nu}^{\,\,\,\,\,\rho}=E_{a}^{\rho}\,\left(D_{\mu}e\right)_{\nu}^{a},$expressed
in terms of components $e_{\mu}^{a}$ and of its inverse $E_{a}^{\rho}$,
besides the usual Christoffel symbols also contains a contorsion term
in $\Gamma_{\left(\mu\nu\right)}^{\,\,\,\,\,\,\,\,\,\,\rho}$, whereas
$\Gamma_{\left[\mu\nu\right]}^{\,\,\,\,\,\,\,\,\,\,\rho}$ are the
components of torsion.

Einstein gravity in d=4 dimensions postulates vanishing torsion $T^{a}=0$
so that $\omega=\omega\left(e\right)$ . This theory can be derived
from the Hilbert action ($G_{N}$is Newton's constant; deSitter space
with nonvanishing positive cosmological constant $\Lambda$ results
from the replacement $R^{ab}\rightarrow R^{ab}-\frac{4}{3}\Lambda e^{a}\wedge e^{b})$

\begin{equation}
L_{\left(H\right)}=\frac{1}{16\pi G_{n}}\int_{\mathcal{M}_{4}}R^{ab}\wedge e^{c}\wedge e^{d}\epsilon_{abcd}+L_{\left(matter\right)},\label{2}\end{equation}

\noindent where a small but definitely nonvanishing value of $\Lambda$
is suggested by recent astronomical observations \cite{Rie-7}. Because
of the {}``Palatini mystery'', independent variation of $\delta\omega$
yields $T^{a}=0,$ whereas $\delta e$ produces the Einstein (or Einstein
- deSitter for$\Lambda\neq0)$ equations.

Instead of working with the metric (\ref{1}) the {}``new'' approaches
\cite{Sen-8} are based upon a gauge field related to $\omega^{ab}$

\begin{equation}
A^{ab}=\frac{1}{2}\left(\omega^{ab}-\frac{\gamma}{2}\epsilon_{\,\,\, cd}^{ab}\omega^{cd}\right).\label{3}\end{equation}

The Barbero-Immirzi parameter $\gamma$ \cite{Bar-9} is an arbitrary
constant. The extension to complex gravity $\left(\gamma=i\right)$
makes $A^{a}$ a self-adjoint field and transforms the Einstein theory
into the one of an $SU$ (\ref{2}) gauge field

\begin{equation}
A_{i}^{\underline{a}}=\epsilon_{\,\,\,\underline{b}\,\underline{c}}^{0\underline{a}}A_{i}^{\underline{b}\,\underline{c}},\label{4}\end{equation}

\noindent where the index $\underline{a}=$1,2,3. This formulation
is the basis of loop quantum gravity and spin foam models (see below).

\section{Observables}

\subsection{Observables in classical GR}

At the classical level the exploration of the global properties of
a certain solution of (\ref{2}), its singularity structure etc.,
is only possible by means of the introduction of an additional test
field, most simply a test particle with action

\begin{eqnarray}
L_{\left(test\right)} & = & -m_{0}\int\left|ds\right|,\nonumber \\
ds^{2} & = & g_{\mu\nu}\left(x\left(\tau\right)\right)\frac{dx^{\mu}}{d\tau}\,\frac{dx^{\nu}}{d\tau},\,\label{5}\end{eqnarray}

\noindent which is another way to incorporate Einstein's old proposal
\cite{Ein-10} of a {}``net of geodesics''. The path $x^{\mu}\left(\tau\right)$
is parameterized by the affine parameter $\tau$ (actually only timelike
or lightlike $ds^{2}\geq0$ describes the paths of a physical particle).

It is not appreciated always that the global properties of a manifold
are \textit{defined} in terms of specific device like (5). Whereas
the usual geodesics derived from (5) depend on $g_{\mu\nu}$ through
the Chistoffel symbols only e.g. in the case of torsion also the contorsion
may contribute ({}``autoparallels'') in the affine connection; spinning
particles {}``feel'' the gravimagnetic effect etc. As a consequence,
when a field dependent transformation of the gravity variables is
performed (e.g. conformal transformation from a {}``Jordan frame''
to an {}``Einstein frame'' in Jordan-Brans-Dicke \cite{Fie-11}
theory) the action of the device must be transformed in the same way.

\subsection{Observables in QFT}

In flat QFT one starts from a Schrödinger equation, dependent on field
operators and, proceeding through Hamiltonian quantization to the
path integral, the experimentally accessible observables are the elements
of the S-matrix, or quantities expressible by those. It should be
recalled that the properly defined renormalized S-matrix element obtains
by amputation of external propagators in the related Green function,
multiplication with polarizations and with a square root of the wave
function renormalization constant, taking the mass-shell limit.

It cannot be emphasized too strongly that in the more general framework
of QFT ordinary quantum mechanics and its Schrödinger equation appear
as the nonrelativistic weak coupling limit of the Bethe-Salpeter equation
of QFT \cite{Sal-12}. Useful notions like eigenvalues of Hermitian
operators, collapse of wave functions etc. are \textit{not} basic
concepts in this more general frame (cf. footnote 2 in ref. \cite{Kum-13}).

In gauge theories one encounters the additional problem of gauge-dependence,
i.e. the dependence on some gauge parameter $\beta$ introduced by
generic gauge fixing. Clearly the S-matrix elements must be and indeed
are \cite{Kum-13} independent of $\beta.$ But other objects, in
particular matrix-elements of gauge invariant operators $\mathcal{O_{A}},$
depend on $\beta.$ In addition, under renormalization they mix with
operators $\widetilde{\mathcal{O}}_{\widetilde{A}}$ of the same {}``twist''
(dimension minus spin) which depend on Faddeev-Popov ghost fields
\cite{Dix-14} and are not gauge-invariant:

\begin{eqnarray}
\mathcal{O}^{\left(ren\right)}= & Z_{AB}\mathcal{O}_{B}+ & Z_{A\widetilde{B}}\widetilde{\mathcal{O}}_{\widetilde{B}}\nonumber \\
\widetilde{\mathcal{O}}^{\left(ren\right)}= &  & Z_{\widetilde{A}\widetilde{B}}\widetilde{\mathcal{O}}_{\widetilde{B}}\label{6}\end{eqnarray}

The contribution of such operators to the S-matrix element (sic!)
of e.g. the scaling limit for deep inelastic scattering \cite{Fri-15}
of leptons on protons \cite{Gro-16} occurs only through the anomalous
dimensions ($\propto\,\partial Z_{AB}/\partial\lambda$ for a regularisation
cut-off $\lambda$). And those objects, also thanks to the triangular
form of (6), turn out to be indeed gauge-independent!

In flat QFT, as well as in QGR, the (gauge invariant) {}``Wilson
loop''

\begin{equation}
W_{\left(\mathcal{C}\right)}=Tr\, P\, exp\,\left(i\oint\limits _{\mathcal{C}}\, A_{\mu}dx^{\mu}\right),\label{7}\end{equation}

\noindent parameterized by a path ordered closed curve $\mathcal{C},$
often is assumed to play an important role. In covariant gauges it
is multiplicatively renormalizable with the renormalization constant
depending on the length of $\mathcal{C},$ the UV cut-off and eventual
cusp-angles in $\mathcal{C}$ \cite{Pol-17}. Still the relation to
experimentally observable quantities (should one simply drop the renormalization
constant or proceed \cite{Kum-13} as for an S-matrix?) is unclear.
Worse, for lightlike axial gauges $\left(n\, A\right)=0\,\left(n^{2}=0\right)$
multiplicative renormalization is not applicable \cite{And-18}. Then,
only for a matrix element of (\ref{7}) between {}``on-shell gluons'',
this type of renormalization is restored. Still the renormalization
constant is different from covariant gauge, except for the anomalous
dimension derived from it (cf. precisely that feature of operators
in deep inelastic scattering of leptons).

In the absence of S-matrix elements, defined as in QFT, nowadays a
broad role is attributed to {}``Dirac observables'', defined as
quantities which commute with the Hamiltonian of the system. (Just
one recent example of this line of argument is ref. \cite{Dit-19}
where also earlier literature is cited extensively). As the example
of the (gauge invariant!) Wilson loops shows, any matrix elements
of the related operators in QFT will be gauge-parameter dependent
and hence useless for a description for a physical phenomena. No convincing
argument is known so far how to extract a \textit{genuine} physical
observable from that.

\section{Traditional and more recent approaches to QGR}

{}``Old'' QGR mostly worked with a separation of the two aspects
of gravity variables by the decomposition of the metric

\begin{equation}
g_{\mu\nu}=g_{\mu\nu}^{\left(0\right)}+h_{\mu\nu},\label{8}\end{equation}

\noindent which consists of a (fixed) classical background $g_{\mu\nu}^{\left(0\right)}$
({}``stage'') upon which small quantum fluctuations $h_{\mu\nu}$
({}``actors'') occur. The {}``observable'' (to be tested by a
classical device) would be the effective matrix $g_{\mu\nu}^{\left(eff\right)}=g_{\mu\nu}^{\left(0\right)}+<h_{\mu\nu}>$.
Starting computations from the action (\ref{2}) one finds that an
ever increasing number of counter-terms is necessary. They are different
from the terms in the Lagrangian $\mathcal{L}=\sqrt{-g}R/\left(16\pi G_{N}\right)$in
(\ref{2}). This is the reason why QGR is called (perturbatively)
{}``nonrenormalizable'' {[}4{]}. Still, at energies $E\ll\left(G_{N}\right)^{-1/2}$,
i.e. much below the Planck mass scale $m_{Pl}\sim10^{19}GeV$, such
calculations can be meaningful in the sense of an {}``effective low
energy field theory'' \cite{Don-20}, irrespective of the fact that
(perhaps by embedding gravity into string theory) by inclusion of
further fields at higher energy scales (Planck scale) QGR may become
renormalizable. Of course, such an approach, even when it is modified
by iterative inclusion of $<h_{\mu\nu}>$into $g_{\mu\nu}^{\left(0\right)}$
etc. - which is quite hopeless technically - , completely misses inherent
background independent effects, i.e. effects when $g_{\mu\nu}^{\left(0\right)}=0$.

One could think also of applying nonperturbative methods developed
in numerical lattice calculations for Quantum Chromodynamics (QCD).
However, there are problems to define the Euclidean path integral
for that, because the Euclidean action is not bounded from below (as
it is the case in QCD) \cite{Gib-21}.

The quantization of gravity which - at least formally - avoids background
dependence is based upon the ADM approach to the Dirac quantization
of the Hamiltonian \cite{Arn-22}. Space-time is foliated by a sequence
of three-dimensional space-like manifolds $\sum_{3}$ upon which the
variables $g_{i\, j}=q_{i\, j}$ and associated canonical momenta
$\pi_{i\, j}$ live. The constraints associated to the further variables
lapse $\left(N_{0}\right)$ and shift $\left(N_{i}\right)$ in the
Hamiltonian density

\begin{equation}
\mathcal{H}=N_{0}H^{0}\left(q,\pi\right)+N_{i}H^{i}\left(q,\pi\right)\label{9}\end{equation}

\noindent are primary ones. The Poisson brackets of the secondary
constraints $H^{\mu}$ closes. $H^{i}$ generates diffeomorphisms
on $\sum_{3}.$ In the quantum versions of (\ref{9}) the solutions
of the Wheeler-deWitt equation involving the Hamiltonian constraint

\begin{equation}
\int\limits _{\sum_{3}}H^{0}\left(q,\,\frac{\delta}{i\delta q}\right)\mid\psi>=0\label{10}\end{equation}

\noindent formally would correspond to a nonperturbative QGR. Apart
form the fact that it is extremely difficult, if not impossible, to
find a general solution to (\ref{10}) there are several basic problems
with a quantum theory based upon that equation (e.g. no Hilbert space
$\mid\psi>$ can be constructed, no preferred time foliation exists
with ensuing inequivalent quantum evolutions \cite{Tor-23}, problems
with usual {}``quantum causality'' exist, the {}``axiom'' of QFT
that fields should commute at space-like distances does not hold etc.).
A restriction to a finite number of degrees of freedom ({}``mini
superspace'') \cite{DeW-24} or to a reduced set of an infinite number
of degrees of freedom (but still less that the original theory - so
called {}``midi superspace'') \cite{Kuc-25} has been found to miss
essential features.

As all physical states $\mid\psi>$ must be annihilated by the constraint
$H^{\mu},$ a naive Schrödinger equation involving the Hamiltonian
constraint $H^{0}$,

\begin{equation}
i\hbar\frac{\partial\mid\psi>}{\partial t}=H^{0}\mid\psi>=0,\label{11}\end{equation}

\noindent cannot contain a time variable. Actually already in ordinary
in quantum mechanics the time variable is something like an outsider
- not associated to any operator, at best defined with reference to
a process which evolves in time. Here the problem even becomes more
serious.

A kind of Schrödinger equation can be produced from (\ref{11}) by
the definition of a {}``time-function'' $T\left(q,\pi,x\right),$
at the price of an even more complicated formalism \cite{Ber-26}
with quite ambiguous results - and the problem persists how to connect
those with {}``genuine'' observables. All these difficulties are
aggravated, when one tries to first eliminate constraints by solving
them explicitly before quantization. In this way clearly part of the
quantum fluctuations are eliminated from the start. As a consequence
different quantum theories, constructed in this way, are not equivalent.

The {}``new'' gravities (loop quantum gravity, spin foam models)
reformulate the quantum theory of space-time by the introduction of
novel variables, based upon the concept of Wilson loops%
\footnote{Recall, however, the serious doubts of a quantum field theory regarding
the use of such variables!%
} (\ref{7}) applied to the gauge field (\ref{4}).

\begin{equation}
U\left(s_{1},\, s_{2}\right)=Tr\, P\, exp\,\left(i\,\int\limits _{s_{1}}^{s_{2}}ds\frac{dx^{i}}{ds}A_{i}\right)\label{12}\end{equation}

\noindent defines a holonomy. It is generalized by inserting further
invariant operators%
\footnote{The same warning should be heeded!%
} at intermediate points between $s_{1}$ and $s_{2}.$ From such holonomies
a spin network can be created which represents space-time (in the
path integral it is dubbed {}``spin foam'').

These approaches claim several successes \cite{Car-2}. Introducing
as a basis diffeomorphism equivalence classes of {}``labeled graphs'',
a finite Hilbert space can be constructed and some solutions of the
Wheeler-deWitt equation (\ref{10}) have been obtained. The methods
introduce a {}``natural'' coarse graining of space-time which implies
a $UV$ cutoff. {}``Small'' gravity around certain states leads
in those cases to corresponding linearized Einstein gravity.

However, despite of very active research in this field a number of
very serious open questions persists: the Hamiltonian constructed
from spin networks does not lead to massless excitations (gravitons)
in the classical limit. The Barbero-Immirzi parameter $\gamma$ has
to be fixed by the requirement of a {}``correct'' Bekenstein-Hawking
entropy for the Black Hole. The most severe problem, however, is the
one of observables. By some researchers in this field it has been
claimed that by {}``proper gauge fixing'' (!) area and volume can
be obtained as quantized {}``observables'', which is a contradiction
in itself from the point of view of QFT. We must emphasize too that
also in an inherently $UV$ regularized theory (finite) renormalization
remains an issue to be dealt with properly. Also the fate of S-matrix
elements, which play such a central role as the proper observables
in QFT, is completely unclear in these setups.

Embedding QGR into (super-)string theory \cite{Pol-27} does not remove
the key problems related to the dual role of the metric. Gravity may
well be a string excitation in a string/brane world of 10 or 11 dimensions,
possibly a finite theory of everything. Nevertheless, at low energies
Einstein gravity (eventually plus an antisymmetric $B-$field) remains
the theory for which computations must be performed.%
\footnote{It should be noted that the now widely confirmed astronomical observations
of a positive cosmological constant \cite{Rie-7} (if it is a constant
and not a {}``quintessence'' field in a theory of the Jordan-Brans-Dicke
type \cite{Fie-11}) precludes immediate application of supersymmetry
(supergravity) in string theory, because only Anti-deSitter space
with $\Lambda<0$ is compatible with supergravity {[}28{]}.%
} Unfortunately, the proper choice (let alone the derivation) of a
string vacuum in our d=4 space-time is an unsolved problem in the
sense that it has too many (billions ?) solutions.

Many other approaches exist, including noncommutative geometry, twistors,
causal sets, 3d approaches, dynamical triangulations, Regge calculus
etc., each of which has certain attractive features and difficulties
(cf. e.g. \cite{Car-2} and refs. therein).

To us all these {}``new'' approaches appear as - very ingenious
- attempts to bypass the technical problems of directly applying standard
QFT to gravity - without a comprehensive solution of the main problems
of QGR being in sight. Thus the main points of a ''minimal'' QFT
for gravity should be based upon {}``proven concepts'' of QFT with
a point of departure characterizing QGR as follows:

(a) QGR at enregies $E\ll m_{Pl}$ is an {}``effective'' low energy
theory and therefore need not be renormalizable QGR to all orders.

(b) QGR is based upon classical Einstein (or Einstein-deSitter) gravity
with usual variables (metric or Cartan variables).

(c) At least the quantization of geometry must be performed in a background
independent (nonperturbative) way.

(d) Absolutely {}``safe'' quantum observables are only the S-matrix
elements $<f\mid S\mid i>$ of QFT where initial state $\mid i>$
and final state $<f\mid$ are defined only when those states are realized
as Fock states of particles in a (at least approximate) flat space
environment. In certain cases it is permissible to employ a semi-classical
approach: expectation values of quantum corrections may be added to
classical geometric variables and a classical computation is then
based on the effective variables obtained in this way.

Clearly item (d) by construction excludes any application to quantum
cosmology, where $\mid i>$ would be the (probably nonexistent) infinite
past before the Big Bang.

Obviously the most difficult issue is (c). We describe in the following
section how gravity models in d=2 (e.g. spherically reduced gravity)
permit a solution of just that crucial point, leading to novel results.

\section{QGR in 1+1 dimension: the {}``Vienna School''}

\subsection{Classical theory: first order formulation}

In the 1990s the interest in dilaton gravity in d=2 was rekindled
by string theory \cite{Man-29}, although in a nonsystematic way many
results were obtained since the 1980s \cite{Bar-30}. For a modern
review on dilaton gravity ref. \cite{Nak-6} may be consulted. The
study of dilaton gravity can be motivated briefly from a purely geometrical
point of view.

The notation of ref. \cite{Gru-5} is used: $e^{a}$ is the zweibein
one-form, $\epsilon=e^{+}\wedge e^{-}$ is the volume two-form. The
one-form $\omega$ represents the spin-connection $\omega_{\,\, b}^{a}=\varepsilon_{\,\, b}^{a}\omega$
with the totally antisymmetric Levi-Civitá symbol $\varepsilon_{ab}\left(\varepsilon_{01}=+1\right).$
With the flat metric $\eta_{ab}$ in light-cone coordinates $\left(\eta_{+-}=1=\eta_{-+},\eta_{++}=0=\eta_{--}\right)$
the torsion 2-form reads $T^{\pm}=\left(d\pm\omega\right)\wedge e^{\pm}$.
The curvature 2-form $R_{\,\, b}^{a}$ can be presented by the 2-form
$R$ defined by $R_{\,\, b}^{a}=\varepsilon_{\,\, b}^{a}R,\,=d\wedge\omega$.
Signs and factors of the Hodge $\star$ operation are defined by $\star\varepsilon=1.$ 

Since the Hilbert action $\int_{\mathcal{M}_{2}}R\propto\left(1-g\right)$
yields just the Euler number for a surface with genus $g$ one has
to generalize it appropriately. The simplest idea is to introduce
a Lagrange multiplier for curvature, X, also known as {}``dilaton
field'', and an arbitrary potential thereof, V(X), in the action
$\int_{\mathcal{M}_{2}}\left(XR+\epsilon V\left(X\right)\right).$In
particular, for $V\propto X$ the Jackiw-Teitelboim model emerges
{[}30{]}. Having introduced curvature it is natural to consider torsion
as well. By analogy the first order gravity action {[}31{]}

\begin{equation}
L^{\left(1\right)}=\int_{\mathcal{M}_{2}}\left(X_{a}T^{a}+XR+\epsilon\mathcal{V}\left(X^{a}X_{a},\, X\right)\right)\label{13}\end{equation}

\noindent can be motivated where $X_{a}$ are the Lagrange multipliers
for torsion. It encompasses essentially all known dilaton theories
in 2d, also known as Generalized Dilaton Theories (GDT). Spherically
reduced gravity (SRG) from d=4 corresponds to $\mathcal{V}=-X^{+}X^{-}/\left(2X\right)-$const. 

Actually (\ref{13}) is classically and quantum mechanically equivalent
to a more familiar expression for a dilaton theory

\begin{equation}
L^{\left(GDT\right)}=\int d^{2}x\sqrt{-g}\left[\frac{R}{2}X-\frac{U}{2}\left(\nabla X\right)^{2}+V\left(X\right)\right],\label{14}\end{equation}

\noindent where the functions $U$ and $V$ coincide with the ones
of a general {}``potential'', quadratic in $X^{a}X_{a}=2X^{+}X^{-},$

\begin{equation}
\mathcal{V}=X^{+}X^{-}U+V.\label{15}\end{equation}

In contrast to (\ref{13}) torsion vanishes in (\ref{14}) and $R$
is the torsionless curvature scalar. The key advantage of the formulation
(\ref{13}) is that it allows an exact (background independent) solution
of the quantum mechanical path integral {[}32{]} - although this solution
has aspects of a {}``topological'' character. Here the fact that
the Hamiltonian of (\ref{13}) leads to a constraint algebra just
like in a nonabelian gauge theory, and the use of a temporal gauge
for the Cartan variables (equivalent to an Eddington-Finkelstein metric
at the level of the metric $g_{\mu\nu}$ in (\ref{14})) are important
ingredients.

On the other hand, the action (\ref{13}) suggested a generalization
to so-called Poisson-Sigma models (PSM \cite{Ike-31})

\begin{equation}
L^{\left(PSM\right)}=\int\left(X^{A}dA_{A}+\frac{1}{2}P^{AB}A_{B}\wedge A_{A}\right).\label{16}\end{equation}

When the Poisson-tensor $P^{AB}\left(X^{C}\right)$ interpreted as
a Schouten bracket $\left\{ X^{A},X^{B}\right\} $, fulfills a genralized
type of Jacoby identity, (\ref{16}) possesses an (on-shell) nonlinear
gauge symmetry of the field $A_{A}$, combined with certain symmetry
transformation in the target space $X^{A}.$ In the mean-time PSM-s
have found applications in string theory \cite{Sch-33},\cite{Sei-34}.

The consideration of graded generalizations of (\ref{16}), i.e. with
anticommuting fields included, in recent years has led to a completely
new approach towards 2d-supergravity \cite{Ert-35} which just for
potentials of type (\ref{15}) (resp. prepotentials derived for them)
allowed to find complete classical solutions (including fermions)
for $N=\left(1,1\right)$ and $N=\left(2,2\right)$ supergravity,
solutions for which the bosonic part alone \cite{Haw-37} was known
previously.

The interaction of 2d dilaton gravity (\ref{14}) with matter, as
exemplified in the special case of SRG, for the first time showed
that the {}``virtual Black Hole'', introduced into an ad hoc manner
before \cite{Haw-37}, naturally appears in a {}``reliable'' S-matrix
situation, namely as an intermediate state in the scattering of spherical
waves \cite{Gru2-38}. Also in the particular case of the {}``stringy''
Black Hole (cf. the last ref. \cite{Man-29}) the quantum correction
to the specific heat could be calculated \cite{Gru3-39} which stabilizes
the (to lowest order) ill-defined value of that quantity.

\section{Conclusion and outlook}

The fundamental challenges of quantum gravity - especially if considered
as a low energy $\left(E\ll m_{Pl}\right)$ theory - are not really
situated in the technical domain. Therefore, they are not likely to
be mitigated in novel formulations which appeared in the last decades.
Rather they are related to the unsolved question how to define genuine
physical observables other than the (gauge-independent) S-matrix elements.
Of course, the latter come into play only in a very restricted domain
of QGR, namely reactions considered in a flat background between infinite
early and late times. For any other situation and in particular for
questions of quantum cosmology, the explicit dependence on gauge parameters
in published results for {}``observables'' signal that we are still
far from understanding basic issues. On the other hand, in a very
restricted domain (2d quantum gravity in a flat background), which
includes spherically reduced Einstein (-deSitter) gravity, many interesting
results can be obtained \cite{Gru-5}. They testify to the fact that
at least in such a situation the uncompromising stance of a quantum
field theory can be maintained to its full extent.

\textbf{\Large Acknowledgement}{\Large \par}

\noindent The author could have not succeeded in pursuing this program
for many years without the large number of collaborators (cf. the
references) of the {}``Vienna Group'' and its eventual extensions.
He thanks in particular the Austrian Science Foundation (FWF) for
financing several projects related to this field, the last one being
P160 30-N08. Among the other forms of support he is especially grateful
to the Austrian Academy of Sciences which in the framework of the
collaboration with the National Academy of Sciences of the Ukraine
co-financed the multilateral research project {}``Quantum Gravity,
Cosmology and Categorification'' and which also supported the travel
expenses to the Kyiv-Sevastopol Meeting 2005. The talk related to
this paper was given by the author at the occasion of the award of
the degree of a honorary doctorate by the National Academy of Sciences
of the Ukraine.

\end{document}